\documentclass[aip,apl, twocolumn, amsmath, amssymb, reprint, groupedaddress]{revtex4-1} 

\usepackage{graphicx}
\usepackage{dcolumn}
\usepackage{bm}
\usepackage{mathrsfs}
\usepackage{amsfonts}
\usepackage{amssymb}

\catcode`\ä = \active
\catcode`\ö = \active
\catcode`\ü = \active
\catcode`\ß = \active
\catcode`\Ä = \active
\catcode`\Ö = \active
\catcode`\Ü = \active
\defä{\"a}
\defö{\"o}
\defü{\"u}
\defÄ{\"A}
\defÖ{\"O}
\defÜ{\"U}
\letß=\ss

\begin{document}

\title{Simple microwave field imaging technique using hot atomic vapor cells}

\author{Pascal Böhi}
\author{Philipp Treutlein}\email[E-mail: ]{philipp.treutlein@unibas.ch}
\affiliation{Departement Physik, Universität Basel, Klingelbergstrasse 82, 4056 Basel, Switzerland}


\begin{abstract}
We demonstrate a simple technique for microwave field imaging using alkali atoms in a vapor cell. The microwave field to be measured drives Rabi oscillations on atomic hyperfine transitions, which are detected in a spatially resolved way using a laser beam and a camera. Our vapor cell geometry enables single-shot recording of two-dimensional microwave field images with $350~\mu$m spatial resolution. Using microfabricated vapor cell arrays, a resolution of a few micrometers seems feasible. All vector components of the microwave magnetic field can be imaged. Our apparatus is simple and compact and does not require cryogenics or ultra-high vacuum. 
\end{abstract}
\maketitle

Integrated microwave circuits are an essential part of modern communication technology and scientific instrumentation. 
For development and testing of such circuits, a technique for high-resolution imaging of microwave field distributions is needed. 
Different methods have been investigated for this purpose,\cite{Sayil05} but a truly
satisfactory standard technique does not exist. 
We recently demonstrated a technique that uses laser-cooled ultracold atoms for microwave field imaging.\cite{Boehi10,BoehiPatentUS} This technique is non-invasive, parallel and therefore fast, offers high spatial and microwave field resolution and allows the
 reconstruction of field amplitudes and phases. 

Although significant progress has been made in simplifying ultracold atom experiments,\cite{Farkas09}
the requirements for ultra-high vacuum and laser cooling
still represent a challenge for microwave field imaging applications in industry.
From a practical point of view it would be very attractive to use thermal atoms in a vapor cell instead, at room temperature or moderately heated. Atomic vapor cells already find widespread use in commercial atomic clocks,\cite{Major10} magnetometers for static and radio-frequency fields,\cite{Budker07,Mikhailov09,Savukov2005,Baranga1998,Young1997} and are investigated as a potential microwave power standard.\cite{Camparo98,Sedlacek12} 
In microwave field imaging, however, high spatial resolution is required, and the fast thermal motion of the atoms presents a serious challenge.


Here we demonstrate a technique for microwave field imaging
that uses an ensemble of near room-temperature atoms in a vapor cell.\cite{BoehiPatentEU} 
Similar to the technique using ultracold atoms,\cite{Boehi10} it relies on position-resolved detection of microwave-driven Rabi oscillations between atomic hyperfine states. 
It provides single-shot 2D imaging, 
reconstruction of amplitudes and phases, and can be made frequency-tunable. 
Since the measured Rabi frequencies only depend on the local microwave field strength and well-known atomic constants, the method is intrinsically calibrated. 
While the proof-of-principle
experiment reported here does not yet reach the same spatial resolution
as the previous technique using ultracold atoms,\cite{Boehi10} we point out how micrometer-scale
spatial resolution can be reached using an array of microfabricated
vapor cells.\cite{Baluktsian10,Hasegawa11,Yang07,Liew04} The experimental setup for our new method is simple, requiring neither cryogenics nor ultra-high vacuum, which makes it promising for the characterization of microwave circuits in real-life applications.

\begin{figure}[tb]
	\centering
		\includegraphics[width=0.65\columnwidth]{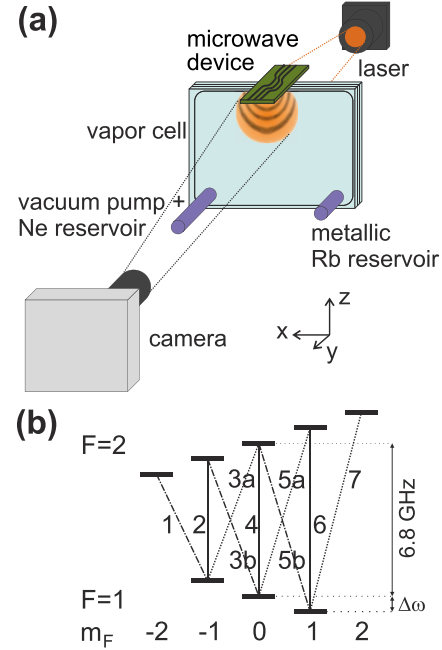}
		\caption{(a) Experimental setup. The microwave device under test is mounted next to an atomic vapor cell, 
		which is connected to a Rb reservoir, a vacuum pump, and a Ne buffer gas reservoir. 
		A laser images the hyperfine state distribution of the $^{87}$Rb atoms in the cell onto a CCD camera. 
		The cell is surrounded by magnetic field coils and sits inside an oven (not shown). 
		(b) Ground-state hyperfine levels $|F,m_F\rangle$ of $^{87}$Rb in a weak static magnetic field. 
		The microwave transitions $i=1...7$ are indicated; transitions 3a+b and 5a+b are degenerate. The laser light selectively images the $F=2$ state.} \label{fig:1}
\end{figure}

The experimental setup is shown in Fig.~\ref{fig:1}a: A vapor cell
is connected to an atomic reservoir, a vacuum pump, and a buffer gas
reservoir. 
In general, any atomic species with microwave transitions that can be read out optically can be used. 
In our specific example we use  $^{87}$Rb atoms in a natural-abundance Rb vapor. 
Neon buffer gas is added to slow down the diffusion of Rb atoms through the cell.\cite{Arditi64} This enhances the spatial resolution of our field imaging method.
The cell has a 2D geometry (much thinner than wide) to allow recording of a 2D image of the microwave field. 
It consists of two quartz glass plates that are epoxy-glued to a 3~mm thick glass frame.
The microwave device to be characterized is mounted at the outside of the cell. It can be moved along the $y$-direction so that 2D images of the microwave field can be taken in any desired cross-section of the device. 
A diode laser provides 780~nm light for optical pumping and for absorption imaging of the atomic hyperfine state distribution onto a charge-coupled device (CCD) camera.
The vapor cell and microwave device are placed inside an oven with two windows for optical access. 
Heating the cell increases the Rb vapor density and thereby enhances laser absorption. 
We point out that our system can also be operated at ambient temperature, albeit at a lower signal-to-noise level. 
A set of Helmholtz coils is used to apply a homogeneous static magnetic field $\mathbf{B}_0$. 
Figure~\ref{fig:1}b shows the ground-state hyperfine structure of $^{87}$Rb in this field. The Zeeman effect results in a splitting of the microwave transition frequencies $\omega_i$ ($i=1...7$) connecting the $m_F$ sublevels of the $F=1$ state to that of the $F=2$ state. For 
$B_0 \ll 0.1$~T, we have $\omega_i=2\pi \times 6.835~\mathrm{GHz}+(i-4)\Delta\omega$, where $\Delta\omega=\mu_B B_0 / 2\hbar$ is the Larmor frequency. 

\begin{figure}[tbp]
\centering
\includegraphics[width=1\columnwidth]{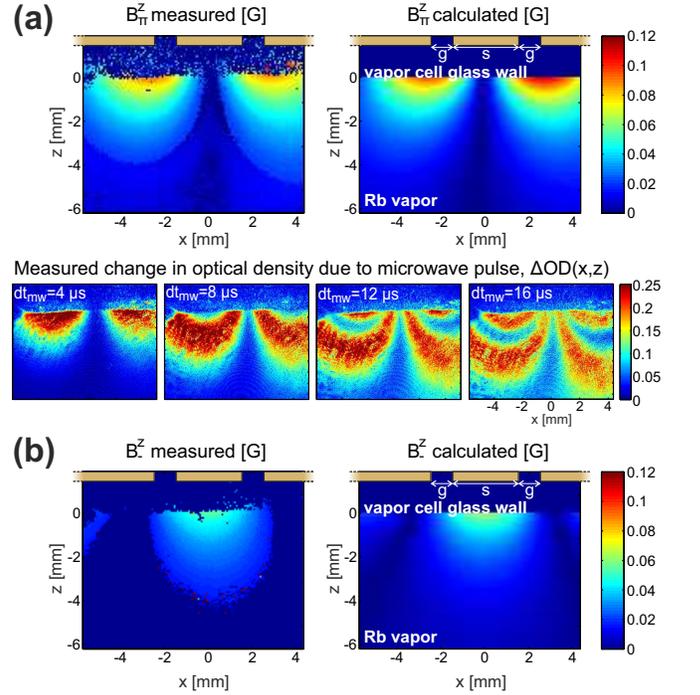}
\caption{Images of two exemplary microwave
magnetic field components near a coplanar waveguide (CPW) and comparison
to a simulation. (a) Top: Spatial distribution of $B_\pi^z$,
i.e.\ the $B_\pi$ polarization component for $\mathbf{B}_0$
pointing along $z$. For this
measurement, we probed the transition $|1,0\rangle \rightarrow |2,0\rangle$ ($i=4$) at $B_0=0.15$~mT. 
On top of the vapor cell, the CPW is indicated ($s=3$~mm, $g=1$~mm, wire thickness not to scale). 
The vapor cell glass wall prevents imaging at distances $<1.5$~mm from the CPW. 
Bottom: Corresponding raw data of the change in optical density $\Delta \mathrm{OD}(x,z)$ for different vaules of $dt_\mathrm{mw}$. This data was used to extract $B_\pi^z$. (b) Distribution
of $B_-^z$ measured on the transition $|1,-1\rangle \rightarrow |2,-2\rangle$ ($i=1$).} \label{fig:2}
\end{figure}

The experimental sequence starts by applying a $100$~ms pulse of laser light to the atoms, which resonantly couples the $F=2$ ground state to the Doppler-broadened excited-state 
manifold of the $^{87}$Rb $D_2$ transition. 
This pulse optically pumps atoms from $F=2$ to  $F=1$, creating a population imbalance between the ground states.\cite{Arditi64,Budker10} 
Next, we turn on the microwave field to be imaged for a duration $dt_\mathrm{mw}$, with frequency $\omega_\mathrm{mw}=\omega_i$ tuned to resonance with one of the 
transitions $i=1,4,7$.
The microwave field drives Rabi oscillations\cite{Budker10} of frequency $\Omega_i(\mathbf{r})$ on the resonant transition, resulting in a modulation of the atomic population in  $F=2$ and a corresponding modulation of the optical density $\mathrm{OD}$ of the vapor of
\[
\Delta \mathrm{OD}(x,z) \propto \sin^2\left[ \tfrac{1}{2} \left| \Omega_i(x,y_0,z) \right| dt_\mathrm{mw} \right],
\]
where $y_0$ is the $y$-position of the cell.
After the microwave pulse, we apply another laser pulse 
and image the light transmitted through the cell onto the camera. 
The duration $dt_\mathrm{im}=10~\mu$s of this imaging pulse is short enough so that optical pumping can be neglected.
We determine $\Delta \mathrm{OD}(x,z) = - \ln \left[ I_\mathrm{mw}(x,z)/I_\mathrm{ref}(x,z) \right]$, where $I_\mathrm{mw}$ ($I_\mathrm{ref}$) is the transmitted  intensity with (without) the microwave pulse applied, see Fig.~\ref{fig:2}a.
We record Rabi oscillations by varying $dt_\mathrm{mw}$ and/or the power $P_\mathrm{mw}$ on the microwave device ($\Omega_i\propto\sqrt{P_\mathrm{mw}}$) and extract $\Omega_i(x,y_0,z)$ for each pixel of the CCD image by a sinusoidal fit.\cite{Boehi10}

The Rabi frequencies $\Omega_i(\mathbf{r})$ are proportional to 
specific components of the microwave magnetic field $\mathscr{B}(\mathbf{r},t)=\tfrac{1}{2}\left[ \mathbf{B}(\mathbf{r})e^{-i\omega_\mathrm{mw}t}+\mathbf{B}^*(\mathbf{r})e^{i\omega_\mathrm{mw}t} \right]$. 
For $^{87}$Rb we have e.g.\
\begin{eqnarray*}
\hbar \Omega_1(\mathbf{r}) & = & -\sqrt{3}\, \mu_B B_-(\mathbf{r})e^{-i\phi_-(\mathbf{r})}, \label{eq:1} \\
\hbar \Omega_4(\mathbf{r}) & = & - \mu_B B_\pi(\mathbf{r})e^{-i\phi_\pi(\mathbf{r})}, \nonumber \\
\hbar \Omega_7(\mathbf{r}) & = & \sqrt{3}\, \mu_B B_+(\mathbf{r})e^{-i\phi_+(\mathbf{r})}. \nonumber
\end{eqnarray*}
Here, $B_\pi$ and $\phi_\pi$ are the real-valued amplitude and phase of the component of $\mathbf{B}$ parallel to $\mathbf{B}_0$, and $\mathbf{B}_+, \phi_+$ ($\mathbf{B}_-, \phi_-$) are the corresponding quantities for the right-handed (left-handed) circular polarization component in the plane perpendicular to $\mathbf{B}_0$. 


The components of the complex microwave magnetic field vector $\mathbf{B}(\mathbf{r})$ can be determined by measuring Rabi frequencies on different transitions.
A single measurement yields the amplitude of one component, but not the phase. To determine amplitudes as well as relative phases of all components, sequential measurements of $B_\pi(\mathbf{r})$, $B_+(\mathbf{r})$, and $B_-(\mathbf{r})$ with the static field $\mathbf{B}_0$ pointing along $x$, $y$, and $z$ are sufficient. From these nine measurements, $\mathbf{B}(\mathbf{r})$ can be reconstructed up to a global phase, as explained in ref.~\cite{Boehi10} It is also possible to determine the spatial dependence of the global phase using an interferometric method.\cite{Boehi10}


To demonstrate our method, we have imaged the microwave field of a simple coplanar waveguide (CPW) with finite ground wires, see Fig.~\ref{fig:2}. 
The CPW is made of $35~\mu$m thick copper wires ($s=3$~mm, $g=1$~mm, ground wires $8$~mm wide) on a $1.5$~mm thick FR-4 substrate and has a characteristic impedance of $70~\Omega$. 
We operate the vapor cell with $p=10$~mbar of Ne buffer gas at a temperature of 
$T\approx 115^{\circ}$C. 
After optical pumping with a laser intensity of $I_0=0.2~\mathrm{mW}/\mathrm{cm}^2$, we measure an optical density $\mathrm{OD} \approx 0.5$. 
The microwave pulse is applied with $P_\mathrm{mw} \approx 5$~W and $dt_\mathrm{mw}$ is varied between 1 and 30~$\mu$s. 
Figure~\ref{fig:2}a (bottom) shows the observed change in optical density $\Delta \mathrm{OD}(x,z)$ for a specific component of the MW field and different values of $dt_\mathrm{mw}$.
The smallest spatial features on the image with $dt_\mathrm{mw}=16~\mu$s have a size of $350~\mu$m (peak-to-valley). 
This is comparable to the calculated r.m.s.\ distance $\Delta x = \sqrt{2 D dt_\mathrm{mw}}=320~\mu$m that a Rb atom diffuses during the microwave pulse. 
Here, $D=D_0 p_0/p$ is the diffusion coefficient for Rb in Ne at pressure $p$, and $D_0=0.31~\mathrm{cm}^2/\mathrm{s}$ is the same quantity at atmospheric pressure $p_0$.\cite{Arditi64}
For larger values of $dt_\mathrm{mw}$, the images start to blur as $\Delta x$ grows and the spatial modulations in $\Delta \mathrm{OD}(x,z)$ get narrower. 
The vapor cell glass wall prevents imaging at distances $<1.5$~mm from the CPW. In an optimized design, the wall thickness would be reduced to allow for imaging at tens of micrometers from the chip.
Two microwave field components reconstructed from such images are shown in Fig.~\ref{fig:2}a+b in comparison with a simulation.\cite{Boehi09}
The simulation agrees well with the measurement if we allow for an 8\% asymmetry between the two ground wire currents. 
While these experiments are a proof-of-principle demonstration using a simple and macroscopic prototype device, much smaller microwave circuits could be characterized with our method as discussed below.

After a few months, the Rb in our cell oxidized. 
We speculate that this is due to the epoxy used to glue our cell (EPO-TEK 353ND). Recent experiments indicate that only certain epoxy glues are compatible with Rb vapor.\cite{PfauPrivComm} 
Another possible explanation is contamination of the buffer gas. The optical density of our cell at $T\approx 115^{\circ}$C drops from $\mathrm{OD}\approx 5$ without buffer gas to $\mathrm{OD}\approx 0.5$ at $p=10$~mbar. Pressure broadening due to buffer gas is known to decrease  optical density,\cite{Rotondaro97} but the decrease we observe is much higher than expected for pure Ne. 
We are confident that these issues can be resolved using a different vapor cell fabrication technique. 
As reported in ref.~\cite{Hasegawa11}, microfabricated alkali vapor cells with Ne buffer gas pressures up to $200$~mbar show very good long-term stability even at elevated temperatures.

One way to increase the spatial resolution of our field imaging method is to use a cell with higher buffer gas pressure to more effectively slow down the Rb motion. 
Neon is particularly well suited as buffer gas because of its small pressure broadening coefficient\cite{Rotondaro97} and small cross section for hyperfine relaxation.\cite{Arditi64}
Microfabricated alkali vapor cells with Ne at $p=100$~mbar have been developed for miniaturized atomic clocks.\cite{Hasegawa11} In such a cell, the atoms diffuse only $\Delta x = 25~\mu$m during $dt_\mathrm{mw}=1~\mu$s. 
This suggests that a transverse spatial resolution of several tens of micrometers can be achieved using buffer gas.
In a cell of $200~\mu$m thickness at $T=100^{\circ}$C, the Rb vapor has $\mathrm{OD}=0.3$, including pressure broadening.\cite{Siddons08,Rotondaro97} 
Hyperfine relaxation is estimated to occur on a timescale of several $\mu$s, dominated by wall collisions.\cite{Arditi64}

Even higher spatial resolution can be obtained by confining the atoms in
an array of micromachined cells fabricated into a substrate.\cite{Baluktsian10,Hasegawa11,Yang07,Liew04}
Each microcell acts as a pixel of the microwave field image.
Using recently developed high-temperature antirelaxation coatings,\cite{Seltzer09}  the hyperfine state of the atoms can survive $\approx 850$ atom-wall collisions at cell temperatures up to $170^\circ$C.
This allows for interaction times up to $dt_\mathrm{mw}\simeq 850 \, d/\bar{v}=13~\mu$s in a cubic cell of side length $d=5~\mu$m  without buffer gas, where the atoms move ballistically with average thermal velocity $\bar{v}=320~\mathrm{m}/\mathrm{s}$ at $T=145^\circ$C. 
Micrometer-scale spatial resolution is thus within reach. 
The optical density is $\mathrm{OD}=0.2$ and Rb spin exchange relaxation is expected to occur on a time scale of $15~\mu$s.\cite{Arditi64}
A small amount of Rb could be deposited in each microcell during fabrication. 
Alternatively, the cells could be connected by thin trenches in the substrate to a single Rb reservoir. 

We now estimate the microwave field sensitivity that can be obtained with such microcells.
Assuming a photon-shot-noise-limited imaging system and $dt_\mathrm{im}=10~\mu$s, the smallest optical density change that can be resolved on a pixel of area $A=d^2$ is $\Delta \mathrm{OD}_\mathrm{min} = \sqrt{\hbar\omega_L / I_0 A \,dt_\mathrm{im}} = 0.02$, where $\omega_L$ is the laser frequency. This implies that a Rabi frequency as small as $\Omega_\mathrm{min} \simeq (2/dt_\mathrm{mw}) \sqrt{\Delta \mathrm{OD}_\mathrm{min} / \mathrm{OD}} =2\pi \times 9$~kHz could  be detected in a single shot.
This corresponds to a microwave magnetic field amplitude of a few hundred nT, which arises e.g.\ at a distance of $5~\mu$m from a microwave guide 
carrying a  signal of $P_\mathrm{mw} \simeq 7$~nW.
In general, a higher spatial resolution implies a lower microwave field resolution, because both the maximal $dt_\mathrm{mw}$ is shorter and the area $A$ over which the image is integrated is smaller.


Microwave field imaging at $6.8$~GHz is useful e.g.\ for the characterization of C-band GaN power amplifiers or scientific applications such as atom chip design.\cite{Boehi09}
However, our method is not restricted to fixed frequency, since the transition frequencies $\omega_i$ can be tuned by means of the static magnetic field $B_0$. Frequencies up to 50~GHz are accessible with standard laboratory fields up to $B_0=1.6$~T.  
For $B_0 > 0.1$~T, the atoms enter the Paschen-Back regime, where the hyperfine transition matrix elements change. 
Nevertheless, a full reconstruction of the microwave field is still possible. At the low-frequency end, the zero-field hyperfine splitting sets a limit. Atoms with smaller hyperfine splitting such as $^{85}$Rb (3.0~GHz) or $^{39}$K (0.5~GHz) give access to low frequencies.
 
The simplicity of our method should allow the construction of integrated
microwave field imaging devices. We envision future devices 
consisting of micromachined vapor cell arrays with integrated
electromagnets to tune the transition frequencies to the desired
values and optical waveguides delivering the light for optical pumping and imaging. 


We thank T.~W.~Hänsch for inspiring discussions, M.~F.~Riedel for help with the experimental setup, A.~Horsley for discussions on vapor cells, and G.-X.~Du for careful reading of the manuscript. This work is supported by the EU project AQUTE and the Swiss National Science Foundation.


\end{document}